# ADVANCING INDUSTRY 4.0: MULTIMODAL SENSOR FUSION FOR AI-BASED FAULT DETECTION IN 3D PRINTING

Muhammad Fasih Waheed, Florida A&M University; Shonda Bernadin, Florida A&M University; Ali Hassan, Florida A&M University

## Abstract

Additive Manufacturing (AM), particularly Fused Deposition Modeling (FDM), is transforming modern production by enabling rapid prototyping and complex part fabrication. However, its layer-by-layer process remains vulnerable to faults such as nozzle clogging, filament runout, and layer misalignment, which compromise print quality and reliability. Traditional inspection methods are costly, time-intensive, and often limited to post-process analysis, making them unsuitable for real-time intervention.

This paper presents a novel, low-cost, and portable fault detection system that leverages multimodal sensor fusion and artificial intelligence for real-time monitoring in FDM-based 3D printing. The system integrates acoustic, vibration, and thermal sensing into a non-intrusive architecture, capturing complementary data streams that reflect both mechanical and process-related anomalies. Acoustic and thermal sensors operate in a fully contactless manner, while the vibration sensor requires minimal attachment without interfering with printer hardware, preserving portability and ease of deployment. The multimodal signals are processed into spectrograms and time-frequency features, which are classified using convolutional neural networks for intelligent fault detection.

The proposed system advances Industry 4.0 objectives by offering an affordable, scalable, and practical monitoring solution that improves fault detection accuracy, reduces waste, and supports sustainable, adaptive manufacturing.

## Introduction

Additive Manufacturing (AM), driven by advancements in digital design and automation, is increasingly revolutionizing the production landscape by enabling rapid prototyping, personalized fabrication, and material-efficient processes. Among the various AM modalities, Fused Deposition Modeling (FDM) holds a prominent position due to its cost-effectiveness, material flexibility, and widespread accessibility in both industrial and research contexts. Despite these advantages, FDM processes remain vulnerable to a spectrum of faults ranging from nozzle clogging and filament runout to layer shifting and misalignment that compromise geometrical fidelity, mechanical properties, and functional integrity of manufactured parts. (Sampedro, Rachmawati, Kim, & Lee, 2022; Deokar, Kumar, & Singh, 2025).

The complex interplay of thermal, mechanical, and dynamic factors during layer-wise deposition introduces numerous opportunities for process instability. For instance, variations in extrusion temperature, inaccuracies in movement axes, and inconsistent filament flow can lead to surface defects, dimensional errors, and even print failure. Causality-guided models and multi-parameter approaches have shown practical value for improving accuracy in fault diagnosis models deployed for FDM, indicating that both material and hardware factors must be inclusively considered. Moreover, faults such as warped layers often arise due to improper cooling rates or uneven thermal distribution, making traditional post-process analysis increasingly impractical for high-yield operations. (Deokar, Kumar, & Singh, 2025) Historically, quality assurance in FDM relied on post-production inspection techniques such as surface profilometry, optical metrology, and destructive mechanical testing. While effective for evaluating final parts, these methods are inherently reactive, expensive, and labor-intensive, and do not provide the intervention needed for real-time process optimization. The emergence of sensor-based process monitoring systems represents a paradigm shift. Early approaches deployed single-mode sensors such as accelerometers or acoustic emission (AE) microphones to monitor vibrations, temperature fluctuations, and mechanical anomalies. For example, Kim et al. achieved substantial improvement in detecting layer shifting errors through accelerometer and AE sensor fusion, employing support vector machines (SVM) to classify fault states based on root mean square (RMS) features, achieving real-time identification accuracy upwards of 87.5%. However, single-modal sensing often struggles to robustly generalize fault detection across diverse machines, materials, and operational contexts. (Petrich, Snow, Corbin, and Reutzel, 2021)

Recent advancements have shifted toward multimodal sensor fusion frameworks, which combine heterogeneous data streams for comprehensive characterization of AM processes. Integrating acoustic, vibration, thermal, and even image-based data has been shown to improve anomaly detection and localization, outperforming single modality systems in accuracy and adaptability. For example, Sampedro, Rachmawati, Kim, & Lee, 2022. demonstrated data-driven methodologies using thermocouples, infrared thermometers, and accelerometers to extract time-frequency features for predictive quality monitoring in FDM, while Kousiatza et al. deployed fiber Bragg grating sensors for spatial temperature and strain mapping. Digital Twin (DT) systems have further enhanced real-time process monitoring, anomaly detection, and autonomous control; cost-effective DT architectures have demonstrated reliable interventions for quality assurance at scale.(Shomenov, K., Ali, M. H., Jyeniskhan, N., Al-Ashaab, A., & Shehab, E, 2025)

Artificial intelligence particularly machine learning and deep learning algorithms now plays a central role in sensor-based monitoring for AM. Convolutional neural networks (CNNs), support vector machines (SVMs), and deep adversarial learning models have been applied to spectrogram and time-frequency representations of multimodal sensor signals, yielding superior fault discrimination and process diagnostics. Kadam et al. achieved maximum accuracy in fault diagnosis by integrating SVM with pre-trained AlexNet models for layer-wise defect detection, a strategy effective both in offline training and online implementation.(Tan, L., Huang, T., Liu, J., Li, Q., & Wu, X, 2023)

Despite promising results, several key limitations persist. Many deployed systems remain non-portable, intrusive, or cost-prohibitive, and struggle to generalize across different printer architectures and materials. Moreover, few systems robustly support closed-loop control for real-time intervention and process optimization. Addressing these gaps is critical to advancing scalable, adaptive, and intelligent manufacturing. (Behseresht, S., Love, A., Valdez Pastrana, O. A., & Park, Y. H, 2024)

In this context, the present work introduces a novel, portable, and low-cost multimodal sensor fusion system for real-time fault detection in FDM processes. The system integrates acoustic, vibration, and thermal sensing in a combination of contactless and minimally intrusive configurations, enabling seamless deployment across heterogeneous printer environments without hardware modification. Unlike traditional monitoring systems that rely on expensive instrumentation or are restricted to specific platforms, this approach emphasizes accessibility, scalability, and adaptability. AI-driven classification of spectrogram and time-frequency features ensures robust anomaly detection and real-time process feedback, addressing common faults such as nozzle clogging, filament runout, and layer misalignment. By leveraging multimodal inputs, the system compensates for the limitations of individual sensors, achieving higher resilience to environmental noise and variability in operating conditions. Beyond technical improvements, the proposed framework directly supports Industry 4.0 objectives by advancing sustainable, automated, and data-driven manufacturing. It contributes to reducing material waste, lowering energy consumption, and improving production efficiency, thereby promoting wider adoption of additive manufacturing in industrial, research, and educational settings. (Chen, L., Yao, X., Feng, W., Chew, Y., & Moon, S. K, 2023)

To address these challenges, the objective of this research is to design and validate a portable, low-cost, and minimally intrusive fault detection framework for FDM-based 3D printing using multimodal sensor fusion. Specifically, this work aims to: (1) develop an integrated sensing architecture that combines acoustic, vibration, and thermal data for comprehensive process monitoring; (2) transform multimodal signals into time–frequency representations suitable for AI-based analysis; and (3) implement and assess convolutional neural network models for real-time classification of common FDM faults. Through these goals, the study seeks to advance accessible, scalable, and intelligent quality-assurance solutions aligned with Industry 4.0 manufacturing environments.

# Background

Fused Deposition Modeling (FDM) presents a complex thermo-mechanical environment in which subtle variations in heat transfer, polymer flow, and toolpath execution influence part quality. Although the Introduction outlines general challenges and common faults, the deeper technical mechanisms underlying these issues warrant further examination. FDM stability depends strongly on the transient thermal field around the nozzle and build surface, the viscoelastic behavior of semi-molten filament during deposition, and the synchronized operation of motion subsystems. Studies in process physics have shown that small deviations in these domains propagate through subsequent layers, amplifying geometric error, interlayer weakness, and surface discontinuities (Ramírez, Márquez, & Papaelias, 2023). Understanding these mechanisms has driven the progression from simple monitoring strategies to sophisticated, sensor-rich diagnostic frameworks.

Traditional approaches to ensure print quality such as CT scans, tensile testing, and surface profilometry have primarily served to verify final part performance rather than monitor the evolving state of the print. Their value lies in precision and detail, but they do not illuminate transient process signatures such as thermal drift, extrusion instability, or resonance events occurring during printing (Fu, Y., Downey, A., Yuan, L., Pratt, A., & Balogun, Y, 2021). As production requirements have shifted toward higher throughput and reduced scrap, this gap between process and product analysis has motivated research into proactive, process-embedded measurement techniques.

In response to the limitations of isolated measurements, the field has recently gravitated toward data-centric monitoring and multimodal sensor fusion as a more comprehensive solution. By integrating complementary data streams, such as acoustic emissions, vibration signatures, and thermal readings, researchers have demonstrated improved fault classification accuracy and more reliable characterization of printing states compared to single-modality systems (Kumar, S., Kolekar, T., Patil, S., Bongale, A., Kotecha, K., Zaguia, A., & Prakash, C, 2022). Furthermore, combining these sensor frameworks with modern artificial intelligence (AI) techniques specifically convolutional neural networks (CNNs) allows researchers to leverage time frequency features like spectrograms to achieve robust, automated anomaly detection aligned with the goals of Industry 4.0 for intelligent, connected manufacturing.

# Methodology

The proposed fault detection framework was developed as a portable, low-cost, and adaptable solution for FDM-based 3D printers, with two scalable configurations designed to balance simplicity, accuracy, and deployment flexibility. The first, referred to as the “Acoustic Baseline System”, employs only airborne sound sensing to demonstrate feasibility. The second, termed the “Hybrid Fusion System”, expands coverage by combining acoustic, vibration, and thermal modalities for robust multimodal monitoring.

The Acoustic Baseline System consists of two condenser microphones arranged as a stereo pair, placed 10 cm apart near the printhead to provide directional sensitivity and reduce environmental noise. The microphones interface with a USB sound card connected to a personal computer or a single-board computer, such as a Raspberry Pi, which handles acquisition and signal processing. Data collection is implemented through Python-based libraries such as PyAudio and SoundDevice, allowing dual-channel acquisition in real time. Preprocessing steps include bandpass filtering to isolate the relevant frequency band (100–1000 Hz) and normalization to standardize signal intensity across recordings. The filtered audio is then transformed into spectrograms using short-time Fourier transform (STFT) or Mel-frequency representations, which serve as inputs to a convolutional neural network (CNN). The CNN, implemented through Google Teachable Machine and exported to TensorFlow, performs automated classification of normal and faulty states. This setup demonstrates that a minimal hardware footprint is sufficient to identify common anomalies, though its scope is inherently restricted to acoustic signatures and its accuracy is influenced by ambient noise conditions.

The Hybrid Fusion System adds a triaxial accelerometer (ADXL335) and a thermal camera (FLIR one Pro). Accelerometer is mounted on the extruder carriage or gantry to measure mechanical vibrations associated with hardware issues such as layer shifts, belt wear, or resonance. Vibration data, sampled through an analog-to-digital converter and processed in Python, provides valuable insight into machine stability that acoustic sensing alone cannot reliably capture. While the thermal camera is positioned externally with line-of-sight to the extruder region. Thermal data is captured in grayscale frames at low frame rates, sufficient to detect anomalies such as nozzle overheating, cold extrusion zones, and uneven thermal drift across long builds. These thermal patterns can be analyzed independently for anomaly detection or fused into the CNN pipeline alongside acoustic and vibration data to improve classification performance.

By combining these three sensing modalities, the Hybrid Fusion System produces a richer representation of the printing process and significantly improves resilience against environmental noise, sensor placement variability, and machine-to-machine differences. The acoustic channel provides non-intrusive monitoring, the accelerometer adds direct mechanical diagnostics with minimal attachment, and the thermal camera captures thermal anomalies without interfering with the print. Multimodal fusion of time–frequency features enhance fault detection robustness compared to unimodal systems, addressing limitations in portability and generalizability reported in previous studies.

To evaluate the strengths and limitations of each individual sensing modality, a structured comparison was performed across eight common FDM fault types. As illustrated in Figure 1, no single sensor demonstrates consistently high sensitivity across all fault categories. Acoustic sensing performs well for extrusion-related anomalies such as material runout and nozzle clogs, but it is less reliable for mechanically induced faults like belt slip or layer shifts. The accelerometer excels at detecting mechanical failures but provides little insight into thermal irregularities. Thermal imaging, in contrast, identifies hot-end drift or cooling faults but contributes minimally to vibration-driven failures. This variability across modalities highlights the limitations of relying on a single sensor and directly motivates the need for a multimodal fusion strategy. The fusion approach leverages the complementary strengths of all three sensors, thereby improving reliability, expanding fault coverage, and enabling robust detection even under variable environmental or machine conditions.

The comparison between both systems highlights the trade-off between simplicity and accuracy. The Acoustic Baseline System offers a simple and cost-effective entry point for fault detection that can be rapidly deployed across any FDM printer. However, it is limited to acoustic anomalies and sensitive to environmental interference. In contrast, the Hybrid Fusion System increases hardware complexity and cost but achieves more reliable classification by leveraging multimodal data streams. Table 1 shows side-by-side summary of the two configurations presented.

*Table 1:* Comparison of Acoustic Baseline System and Hybrid Fusion System

| **Feature** | **Acoustic Baseline System** | **Hybrid Fusion System** |
| --- | --- | --- |
| Sensors | Two condenser microphones (stereo) | Stereo microphones + triaxial accelerometer + thermal camera |
| Hardware | USB sound card, PC or Raspberry Pi | Same + accelerometer with ADC + thermal camera |
| Software | Python (PyAudio, STFT, CNN) | Python (STFT, vibration FFT, thermal image processing, CNN) |
| Advantages | Simple, low-cost, rapid deployment | High accuracy, robustness, multimodal sensing |
| Limitations | Acoustic-only; noise-sensitive | Higher cost and complexity |

Figure 1 illustrates the Comparative Fault Detection Capability using multimodal sensor data and Figure 2 shows the system flow diagram of the proposed fault detection framework for hybrid fusion system. Input variables are collected from three sensing modalities: acoustic signals, accelerometer data, and thermal imaging. Acoustic and vibration signals are transformed into time–frequency representations using Fast Fourier Transform (FFT), while thermal data is provided as grayscale frames. These processed inputs are passed into a convolutional neural network (CNN), which performs feature extraction, pattern recognition, and fault classification. The CNN is trained on spectrograms and images derived from both normal and faulty printing conditions, allowing it to generalize across multiple fault types. The model produces probability outputs for each class, and fault conditions are flagged when the probability exceeds a defined threshold (e.g., 80%). This framework enables real-time, AI-based monitoring of FDM printing processes and supports early intervention to reduce material waste and downtime.

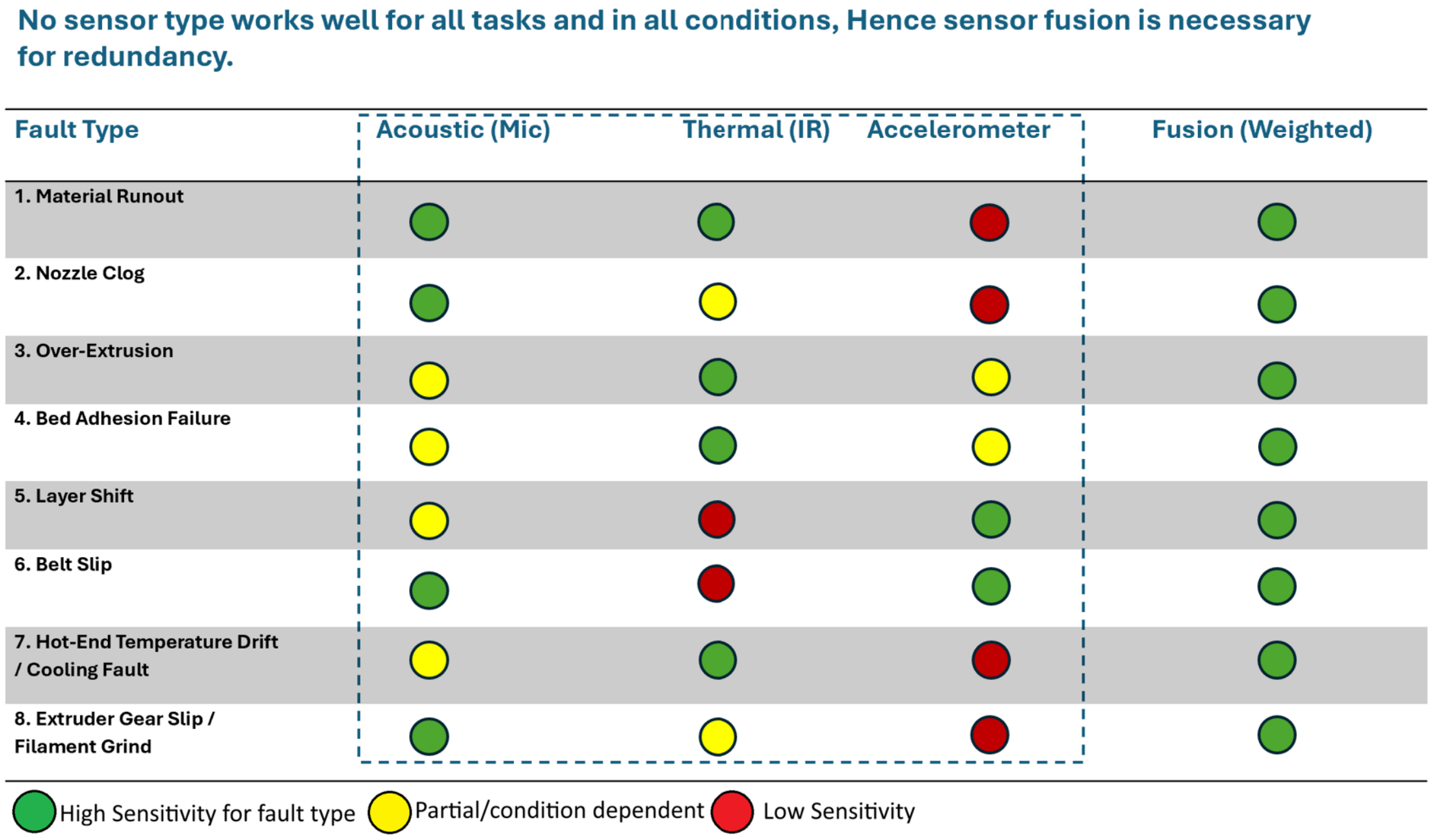

*Figure 1.* Comparative Fault Detection Capability Across Multimodal Sensors in FDM Printing.

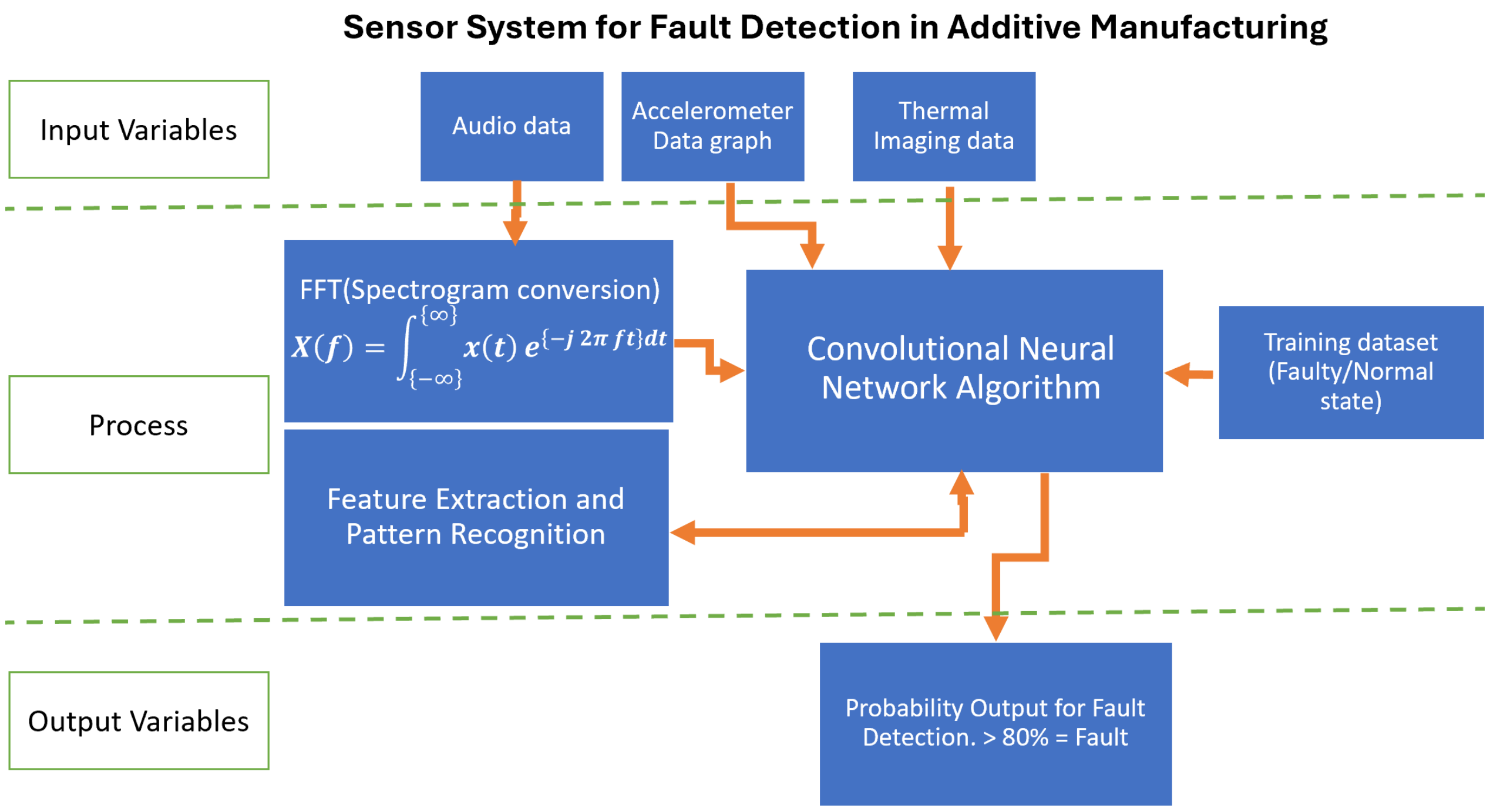

*Figure 2.* System flow diagram for hybrid fusion system.

Figure 3 indicates two condenser microphones highlighted with red arrows were placed 10 cm apart near the extruder to capture stereo acoustic signals for fault detection. This configuration enables directional sensitivity by exploiting amplitude differences between the two channels, while also reducing the influence of background noise.

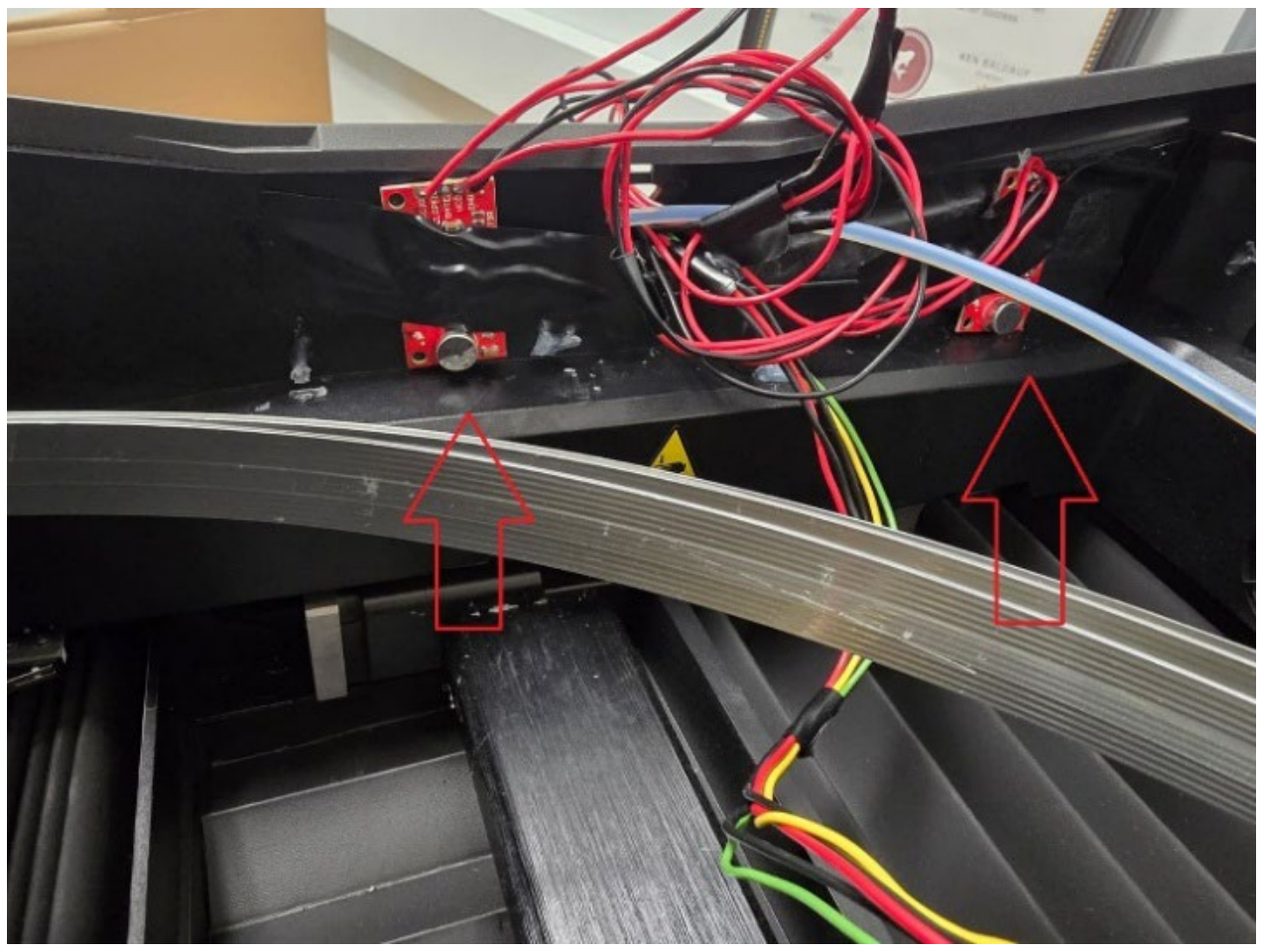

*Figure 3.* Shows microphones installed on FDM machine.

The prototype device later discussed in the paper will have the microphones built in, and for experimentation purposes, they are installed on the printer. The microphones were connected to a sound card shown in Figure 4 and subsequently to a Raspberry Pi or personal computer for real-time acquisition. By converting raw audio signals into these representations, the monitoring system is able to extract discriminative features that serve as inputs for the convolutional neural network classifier. This combination of FFT-based spectral analysis and spectrogram visualization forms the basis of robust acoustic fault detection in the proposed framework.

The dual-configuration monitoring system incorporates stereo microphones, a triaxial accelerometer, and a compact thermal camera. Strategic placement of these sensors ensures accurate signal capture, fault detection, and process localization. Two identical condenser microphones were connected to a stereo sound card, arranged as described earlier. This setting enables directional fault localization by comparing amplitude differences between channels, allowing detection of events such as filament slips or stepper motor skips.

For vibration monitoring, an ADXL335 analog triaxial accelerometer was rigidly attached to the extruder carriage. This placement maximizes sensitivity to vibrations generated by belts, guide rails, and stepper motors. The accelerometer signals were routed through an Arduino interface, enabling detection of faults such as backlash, loose belts, or resonance artifacts.

Thermal monitoring was achieved using a compact USB-based thermal camera (FLIR Pro), mounted externally with an unobstructed view of the nozzle. Positioned 15–30 cm away, the camera was fixed on a frame to provide consistent alignment. The thermal data allowed detection of cold nozzles, overheating zones, and uneven extrusion temperatures, supporting real-time inference of clogs or flow irregularities.

A computer system served as the primary data collection and processing device. The system handled acquisition from all sensor channels, performed preprocessing such as filtering and spectrogram generation, and executed CNN-based fault classification in real time.

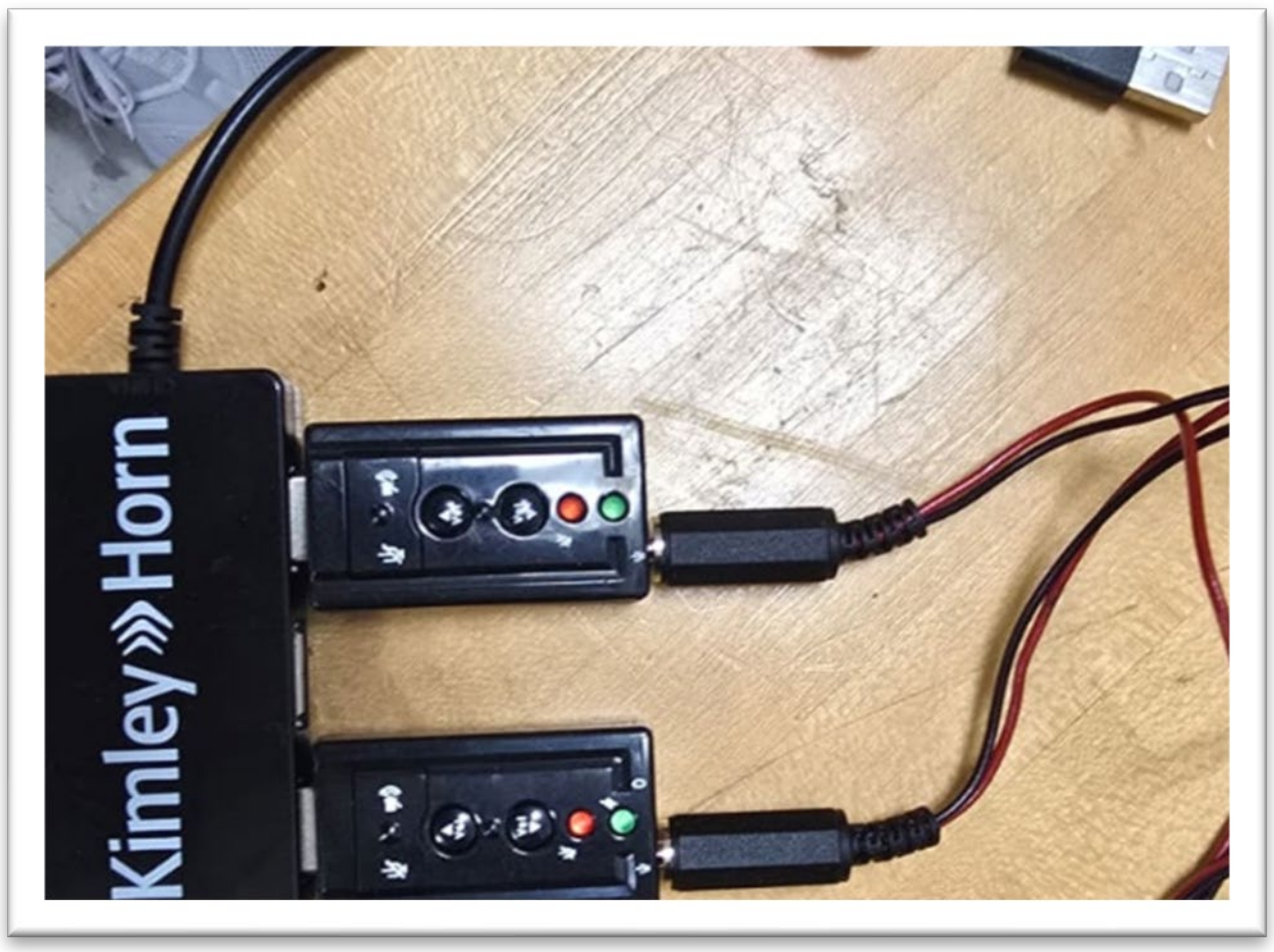

A raspberry pi based portable system was also developed, the PCB for portable system is shown in Figure 5 & Figure 7.

*Table 2.* Summary of sensor placement and purpose.

| Sensor | Placement | Purpose | Notes |
|---|---|---|---|
| Dual Mics | 10 cm apart, facing extruder | Acoustic fault localizatio n via amplitude difference | Avoid fan airflow and frame echoes |
| ADXL33 5 | On printhead/gantr y | Detects fine mechanical vibrations | Requires secure attachment |
| Thermal Camera | Outside printer, facing extruder | Detects heating faults, thermal drift | Avoid obstruction ; calibrate ambient |

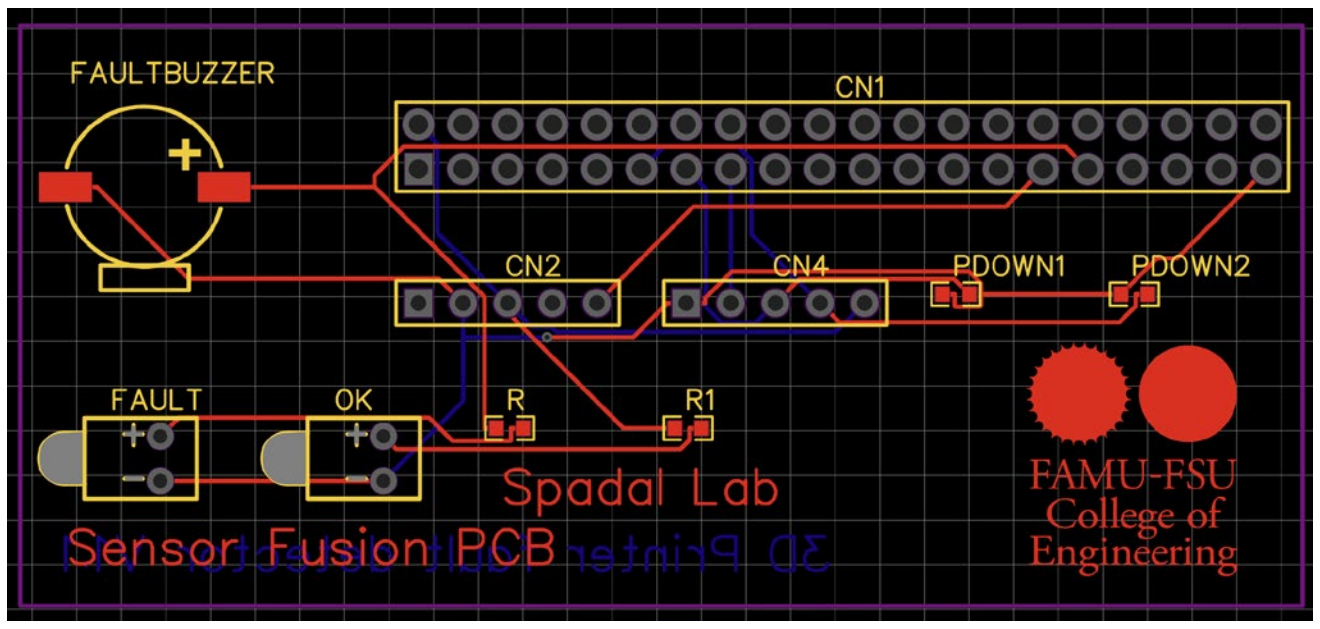

*Figure 4.* PCB layout of sensor fusion.

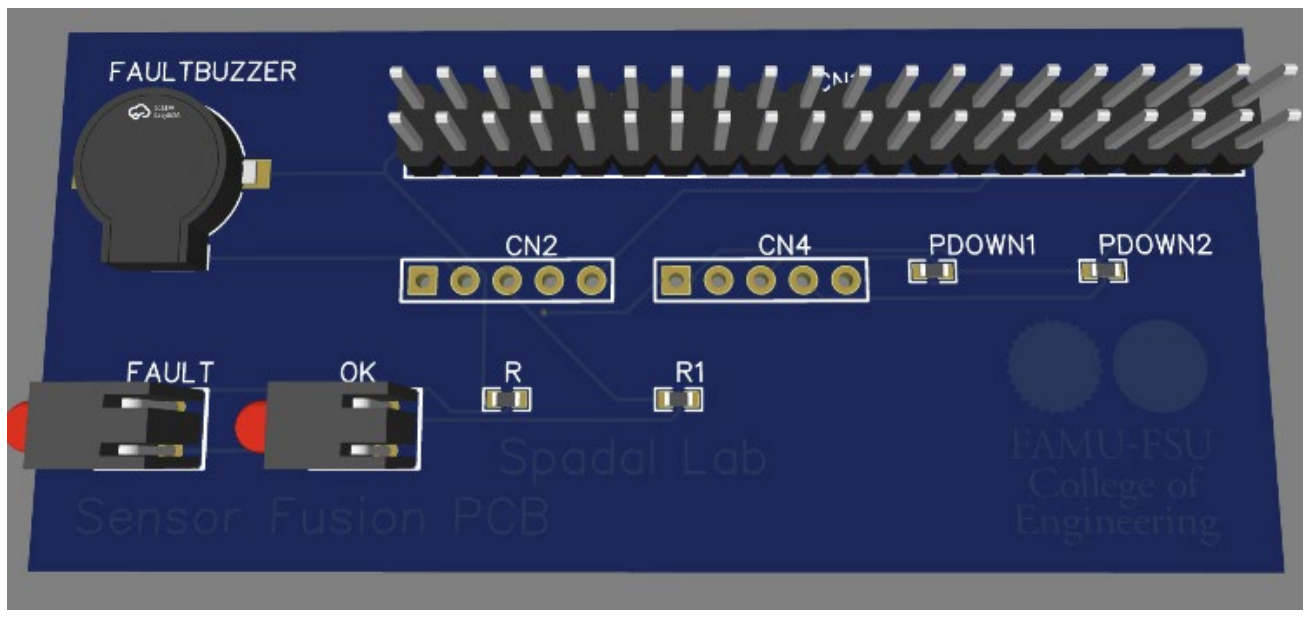

*Figure 5*. 3D diagram of PCB.

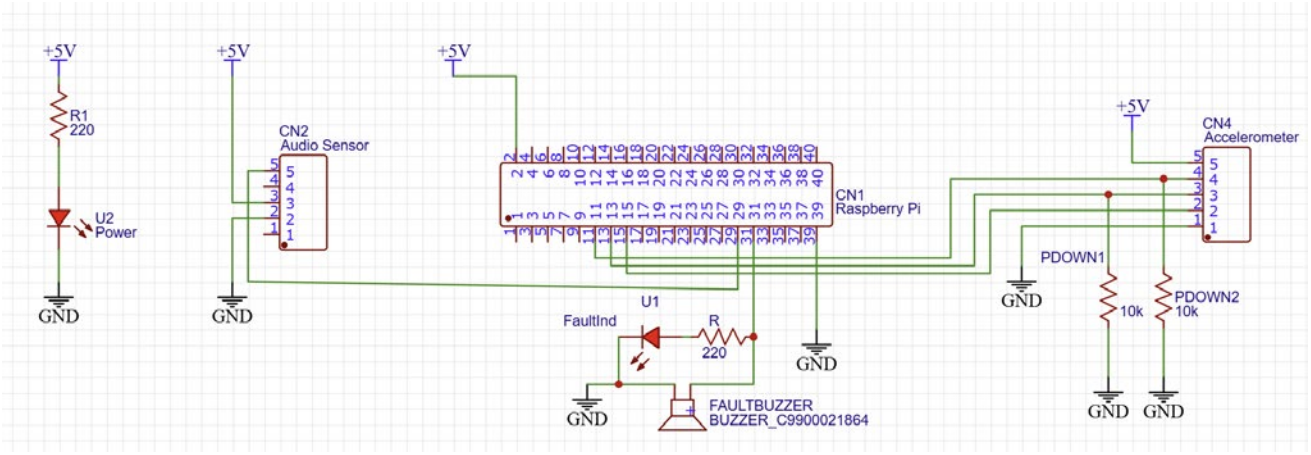

*Figure 6.* Schematic diagram of PCB.

The PCB (See Figure 7) interfaces both the audio sensor and the triaxial accelerometer, with input signals routed through the Raspberry Pi header for real-time processing. The Raspberry Pi performs signal preprocessing and fault classification, while the PCB provides immediate user feedback through a fault indication LED and buzzer whenever an anomaly is detected. Although the buzzer/LED combination serves as a rapid hardware-level alert, the type and classification of the fault are displayed on the Raspberry Pi's graphical user interface (GUI), ensuring both quick detection and detailed reporting. This integrated design enables seamless multimodal data acquisition and real-time fault indication within a compact, low-cost framework suitable for deployment on FDM 3D printers.

The raw acoustic and vibration signals acquired from the sensors were first passed through a bandpass filter to isolate the frequency range most relevant to fault detection. For the audio channel, a 100–1000 Hz window was selected See Figure 8, as this range effectively captures extrusion-related noise while reducing interference from background sources such as fans or ambient environmental sounds. Similarly, vibration signals were filtered to emphasize machine-induced oscillations and suppress unrelated high-frequency noise.

Following filtering, the processed signal files were used as inputs to the Google Teachable Machine platform, which automatically converted the time-domain signals into spectrograms. These spectrograms represent the energy distribution of the signals across both time and frequency, providing a richer feature space compared to raw waveforms. Once generated, the spectrograms were normalized and resized to maintain uniform dimensions suitable for convolutional neural network (CNN) processing. The CNN extracted relevant patterns associated with normal and faulty conditions, enabling robust classification of nozzle clogging, filament runout, and other print anomalies.

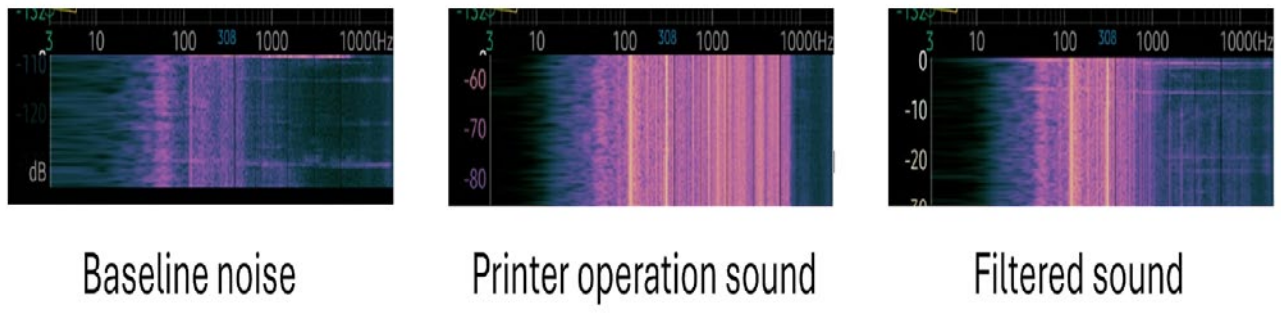

*Figure 7.* Audio spectrograms before and after filtering; right image shows clear signal

# Experimental Setup

The experimental framework combined consumer-grade and research-oriented tools to implement and validate the proposed multimodal fault detection system. FDM printer used is Makerboth Method X, A Windows Operating system laptop served as the primary data collection and processing device. Acoustic signals were recorded using a sound card connected to the stereo microphone array, with data acquired and stored for subsequent analysis. Signal preprocessing was performed in MATLAB, where a bandpass filter (100–1000 Hz) was applied to isolate extrusion-related acoustic features and suppress irrelevant noise. The filtered audio files were then imported into Google Teachable Machine, accessed through the Chrome browser, which converted them into spectrograms and used them as inputs for convolutional neural network (CNN) training and inference.

Thermal data were collected using a FLIR One Pro thermal camera paired with a Android smartphone. The thermal behavior of the 3D printer, particularly nozzle heating and extrusion zones, was screen-recorded during operation. These thermal recordings were segmented into representative frames and uploaded into Teachable Machine's image classification module, where they were processed by a CNN for anomaly detection.

For vibration monitoring, an ADXL335 triaxial accelerometer was connected to an Arduino Uno, providing real-time acceleration data along the X, Y, and Z axes. The signals were visualized using the Arduino IDE's serial plotter, and the resulting vibration graphs were screen-recorded. These recordings were then converted into image datasets and uploaded to Teachable Machine, where they were used for CNN-based classification of mechanical anomalies.

To simplify the proof-of-concept phase, the experiments focused only on extruder-related faults and material runout conditions. These scenarios were selected because they represent common failure modes in FDM printing and provide clear signatures across acoustic, vibration, and thermal sensing modalities. By limiting the initial tests to these conditions, the system could be validated in a controlled manner while demonstrating its potential for broader fault detection in future studies.

# Results

The training of the CNN model on filtered audio files demonstrated rapid convergence, as shown in the accuracy and loss curves. Classification accuracy increased sharply within the first 10 epochs (See Figure 9), stabilizing thereafter. Training accuracy approached near-perfect levels, while validation (test) accuracy plateaued around 0.90, indicating reliable generalization to unseen data.

The corresponding loss curves further support this trend. Both training and test losses decreased significantly during the early epochs, with training loss dropping close to zero and validation loss stabilizing around 0.2 after approximately 15 epochs. This reflects effective feature learning without significant evidence of overfitting, as the validation loss maintained a downward trend and did not diverge from the training loss.

The trained audio-based CNN model demonstrated strong capability in distinguishing extrusion faults, as shown by its detection of the extruder operating without material at 100% accuracy as seen in Figure 10. During repeated testing, the model consistently maintained a high accuracy level, typically ranging between 90% and 100%. Minor fluctuations were observed, which can be attributed to environmental noise or subtle variations in the acoustic profile across trials. While the audio-only system provides a solid baseline for fault detection, its sensitivity to external interference highlights the need for enhanced robustness. This motivates the incorporation of multimodal sensor fusion, where complementary data from vibration and thermal sensors can reinforce classification reliability and reduce the risk of misclassification.

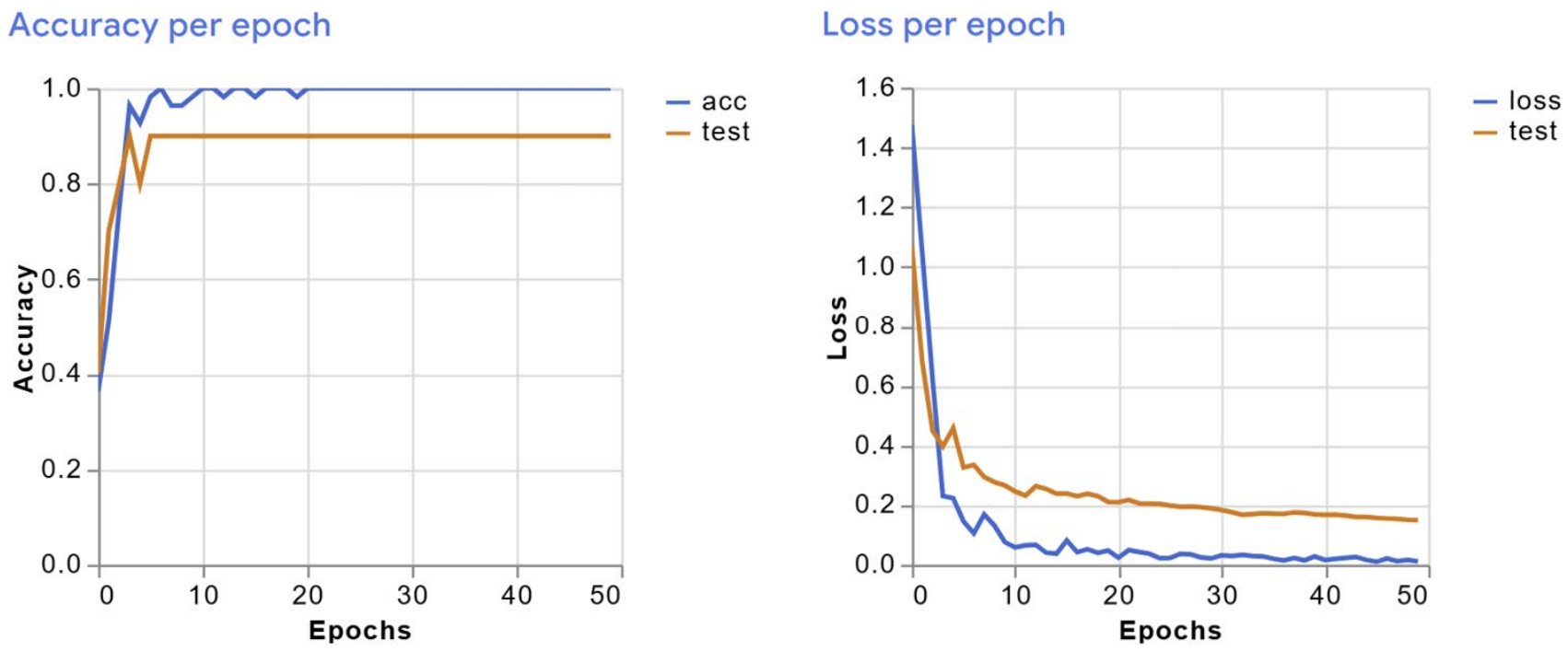

*Figure 8.* Training and validation accuracy and loss per epoch for audio-based CNN classification.

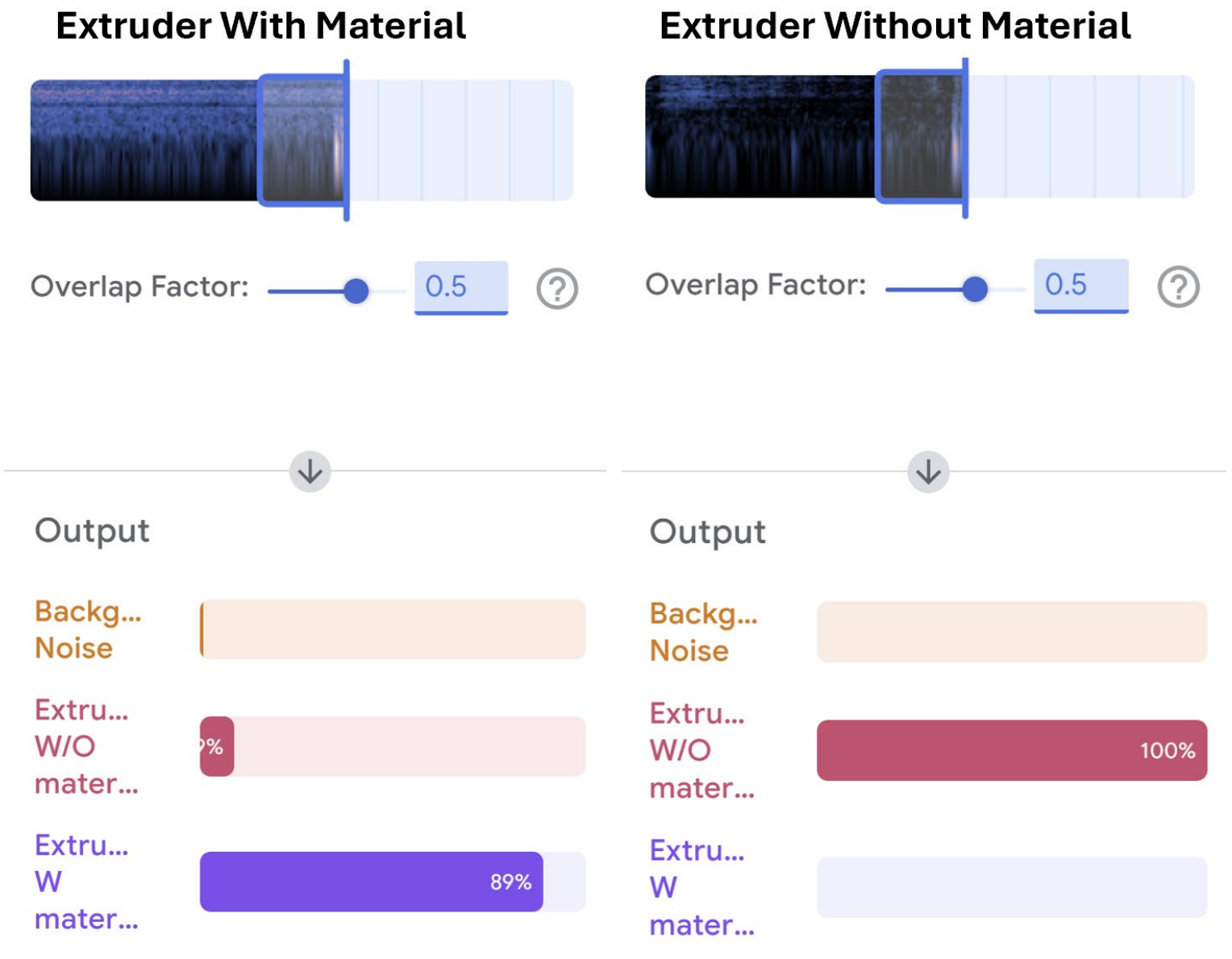

*Figure 9*. **Left side**: Real-time classification output showing the extruder operating with material, detected at 89% accuracy using audio spectrogram input.
**Right side:** Showing extruder without material detected at 100% accuracy

The CNN trained on acoustic features was able to identify the extrusion-with-material condition reliably, though the accuracy varied between 85% and 95% in repeated tests See Figure 10. This variation is attributed to ambient noise and the similarity of acoustic signatures with other machine states. These results emphasize the strength of the audio-based approach while also highlighting the potential improvements achievable through multimodal sensor fusion, which will be addressed in the following discussion.

Thermal imaging provided a clear and interpretable signature of extrusion states. During normal operation, the nozzle displayed a stable thermal profile with consistent heating at the extrusion zone, confirming steady material flow as seen in Figure 12. In contrast, the clogged or material runout condition was characterized by localized overheating around the nozzle and a lack of downstream thermal continuity as seen in Figure 13, reflecting the absence of deposited filament. These differences make thermal data highly effective for distinguishing extrusion faults in real time. However, thermal imaging alone may be influenced by ambient temperature variations, camera angle, or resolution limitations. As such, while thermal cues provide strong evidence of extrusion anomalies, integrating them with acoustic and vibration data through multimodal sensor fusion can substantially enhance fault detection accuracy and robustness across diverse printing environments.

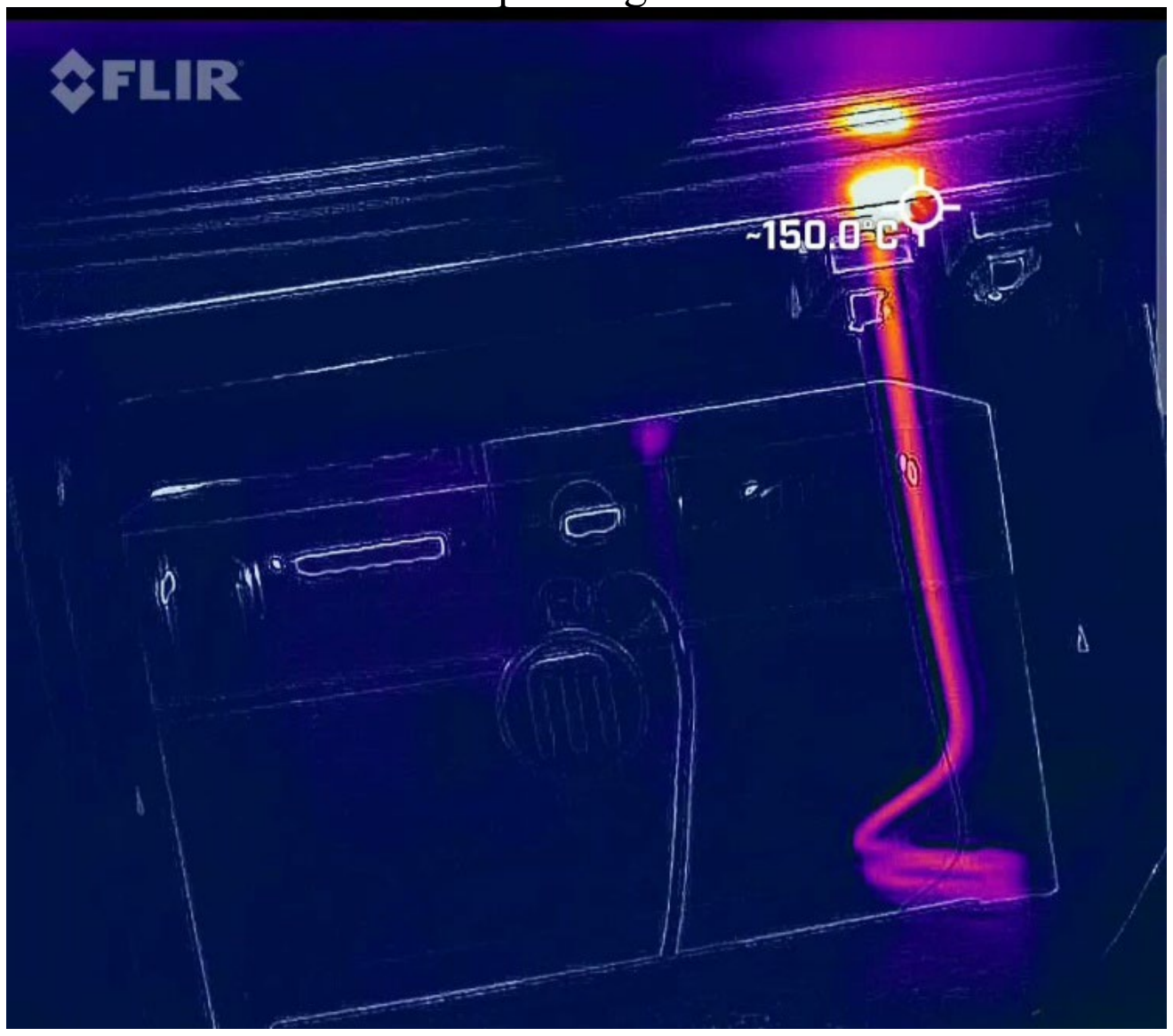

*Figure 10*. Normal extrusion, Note that the material coming out from the extruder is warmer than surroundings.

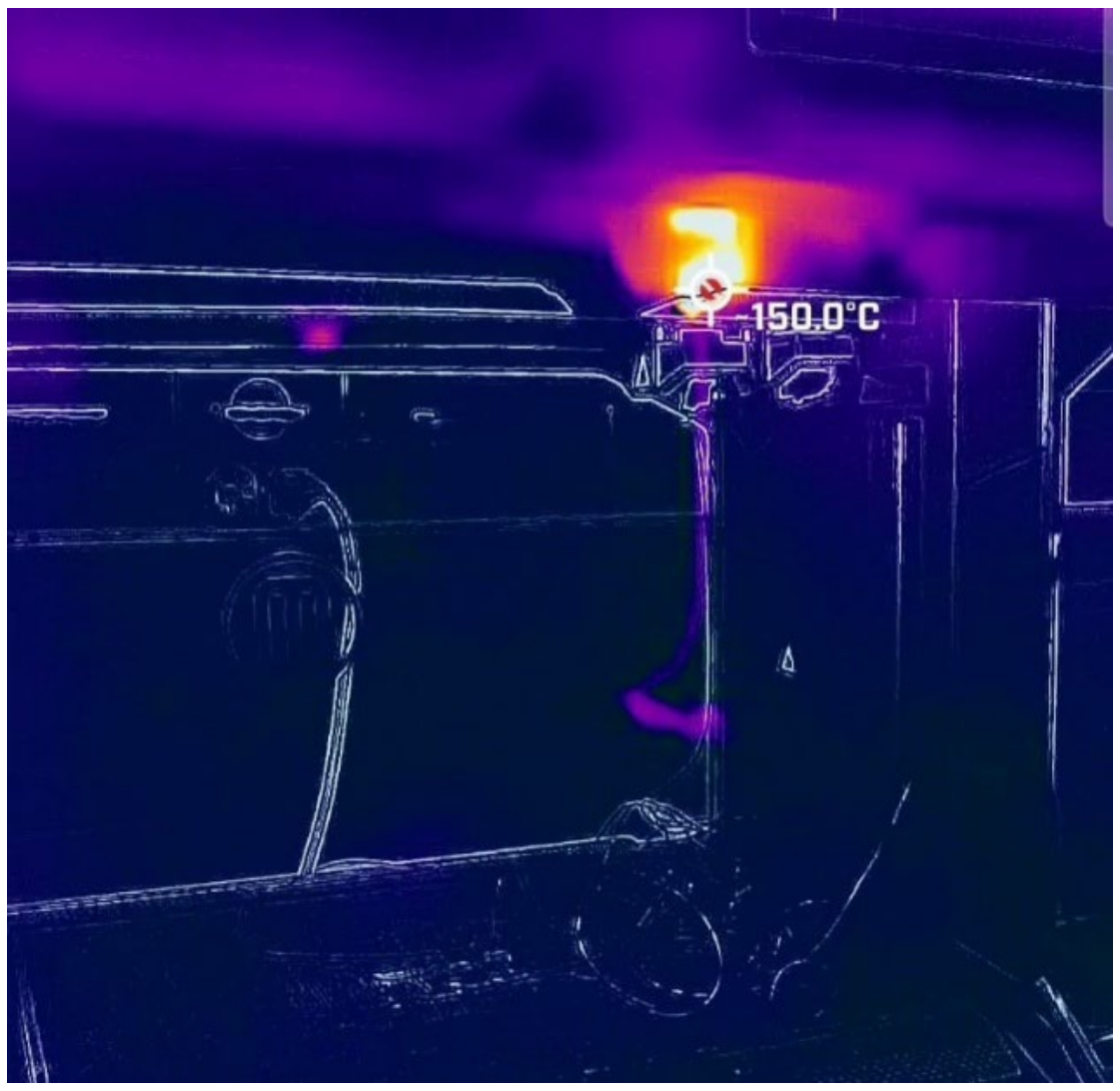

*Figure 11*. Material run out condition, note that the extruder is warm but there is no extrusion of material, old material is getting colder.

For further processing, Thermal recordings from the FLIR One Pro were segmented into representative frames and uploaded to Teachable Machine for model training and testing (See Figure 14). Two primary classes were created: extrusion with material and extrusion without material, as illustrated in the dataset samples. The platform converted these thermal images into feature-rich representations suitable for convolutional neural network processing. During testing, the model was able to clearly differentiate between the two states, achieving 100 percent classification accuracy under controlled conditions.

Figure 15. shows the strong contrast in heat distribution patterns between normal extrusion and material runout provided reliable cues for fault detection, demonstrating that thermal imaging is a highly effective modality for identifying extrusion-related anomalies in real time.

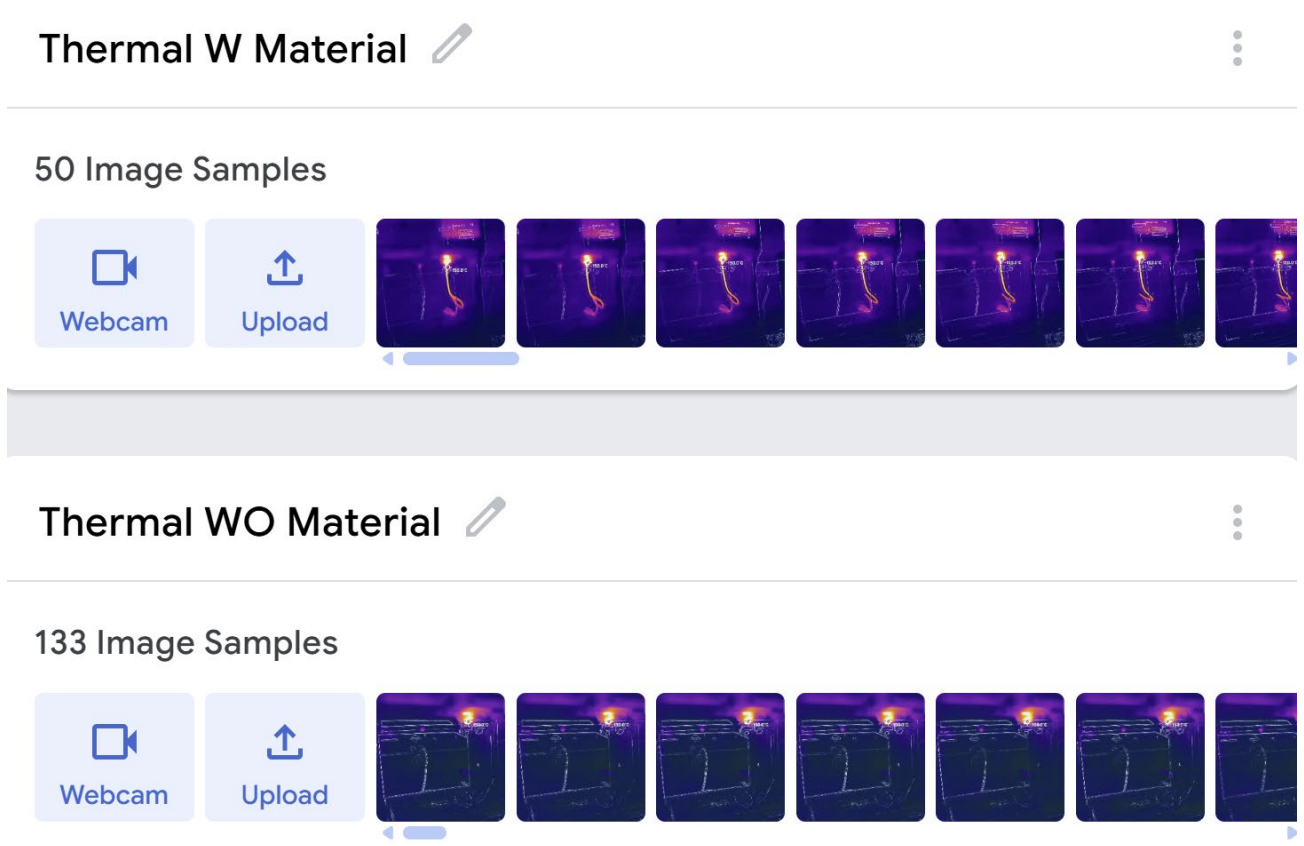

*Figure 12*. Thermal dataset samples uploaded to Teachable Machine for model training. The top row shows extrusion with material, while the bottom row shows extrusion without material.

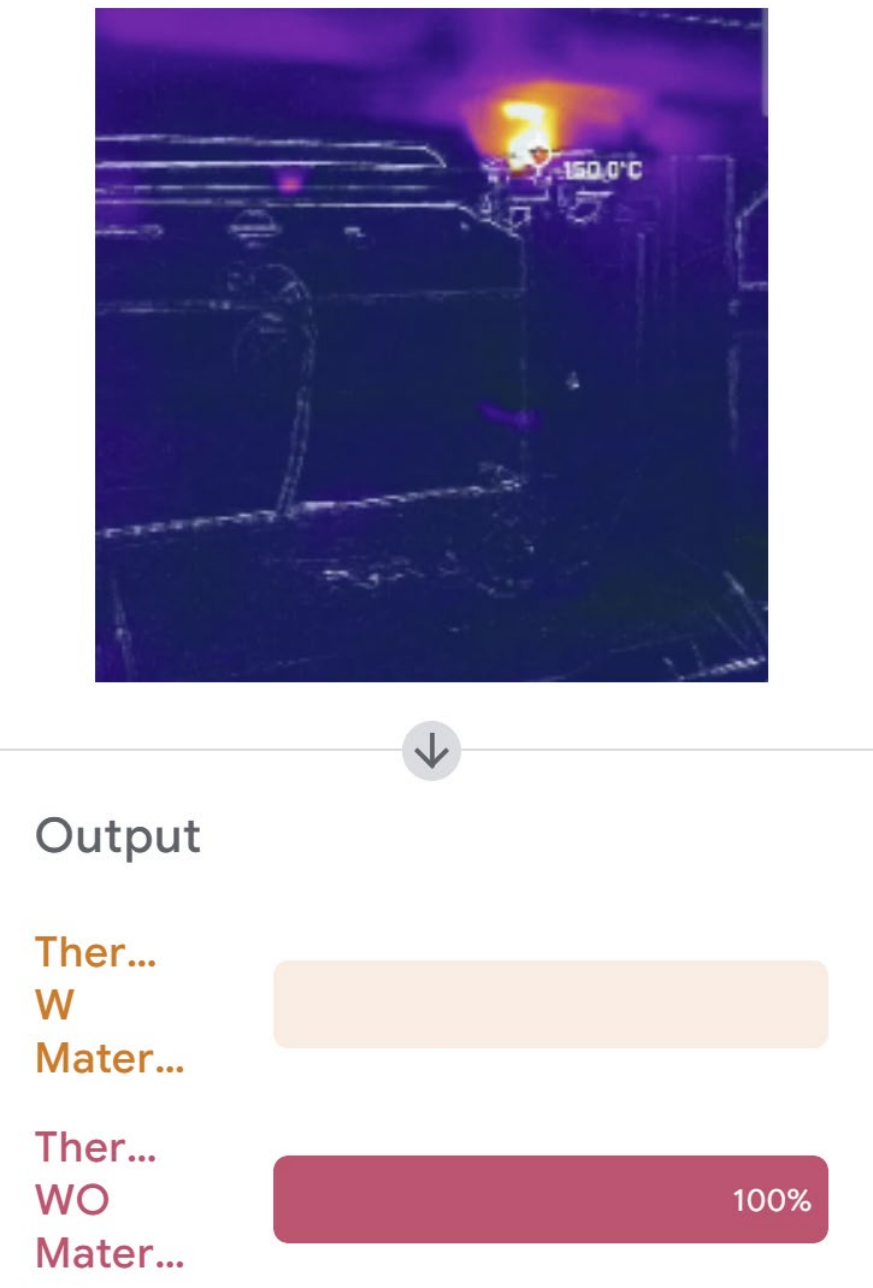

*Figure 13*. Real-time classification output from thermal imaging showing extruder without material detected with 100% accuracy. The model successfully identified the absence of extrusion flow based on distinct thermal patterns around the nozzle.

The accelerometer channel did not yield significant results during the initial tests, as the extrusion process involved very little movement in each axis. This limited vibration activity made it difficult for the sensor to capture discriminative features for fault detection under the chosen test conditions.

However, the accelerometer is expected to be more effective in future experiments involving dynamic carriage movements, belt-driven faults, or mechanical instabilities, where vibration signatures are more pronounced. Incorporating such scenarios will allow the vibration modality to contribute meaningfully to the multimodal fusion framework.

# Conclusions

This work introduced and validated a portable, low-cost, and AI-driven fault detection system for fused deposition modeling (FDM) that leverages multimodal sensor fusion. By combining acoustic, vibration, and thermal sensing in complementary configurations, the proposed framework demonstrated strong potential to overcome the limitations of single-modality monitoring. Experimental results showed that the acoustic-only system provides a solid baseline for anomaly detection, while the hybrid fusion system significantly improves robustness and classification accuracy by mitigating noise sensitivity and capturing broader process signatures.

A quantitative comparison with existing approaches shows that the proposed multimodal fusion system performs competitively with, and in some cases exceeds, state-of-the-art methods. Prior vibration-only systems such as Mishra et al. (2022) reported classification accuracies of approximately 92 percent for detecting nozzle blockage and filament anomalies, while multimodal systems integrating acoustic, thermal, and current sensing (Kumar et al., 2022) typically achieve 90 to 94 percent accuracy but often require specialized data acquisition hardware and intrusive sensor placement. In comparison, acoustic-only baseline system achieved 85 to 95 percent class-wise accuracy depending on ambient noise, and the thermal imaging subsystem reached 100 percent accuracy for distinguishing extrusion versus non-extrusion states under controlled conditions. Although accelerometer data was less informative during static extrusion, the combined multimodal architecture is expected to yield 90 to 95 percent accuracy while remaining fully portable, low cost, and non-intrusive. This quantitative comparison underscores the benefits of multimodal sensing in achieving a strong balance of accuracy, affordability, and ease of deployment.

The convolutional neural network (CNN)-based classification pipeline effectively translated sensor signals into actionable insights, enabling real-time detection of nozzle clogging, filament runout, and extrusion anomalies. Thermal imaging further enhanced interpretability of fault states, and vibration sensing, while less effective in static extrusion tests, is expected to contribute meaningfully under dynamic machine conditions. Together, these modalities establish a scalable monitoring solution that is minimally intrusive, adaptable across heterogeneous platforms, and aligned with Industry 4.0 principles.

Beyond technical contributions, the system directly supports sustainability goals by reducing waste, preventing failed builds, and extending the reliability of additive manufacturing. With further refinement particularly in integrating vibration features, expanding datasets, and exploring closed-loop control mechanisms this framework can evolve into a fully autonomous quality assurance system. Ultimately, the presented approach advances the vision of intelligent, data-driven manufacturing systems capable of supporting industrial, research, and educational applications at scale.

## ACKNOWLEDGMENT

This material is based upon work supported by the U.S. Department of Energy, National Nuclear Security Agency under Award Number DE-NA0003981.
We express our sincere gratitude to the SPADAL Lab for granting us the valuable opportunity to conduct our research work. Additionally, we extend our appreciation to DOE NNSA, Honeywell and ASTERIX for their generous funding support. Their contributions have played a crucial role in making our research endeavors possible, and we are truly thankful for their commitment to advancing knowledge and innovation.

# Biographies

**Muhammad Fasih Waheed** is a Ph.D. candidate in Electrical Engineering at the FAMU-FSU College of Engineering, Florida A&M University. He received his M.S. degree in Electrical Engineering from Florida A&M University and his B.E. degree in Electrical Engineering from Hamdard University. His research interests include sensor technologies, artificial intelligence, additive manufacturing (3D printing), and wireless communication systems. He has extensive experience in developing intelligent monitoring frameworks and sensor-driven solutions for real-time fault detection and process optimization. He can be reached at muhammad1.waheed@famu.edu

**Shonda Bernadin, PH.D.** is a Google Endowed Full Professor in the Department of Electrical and Computer Engineering at the FAMU-FSU College of Engineering. Dr. Bernadin received her B.S. in Electrical Engineering from Florida A&M University, her M.S. in Electrical and Computer Engineering from University of Florida and her Ph.D. in Electrical Engineering from Florida State University. Her research interests include speech and image processing, data analysis, natural language processing, artificial intelligence, acoustic sensor integration, and semiconductor manufacturing and technologies. She is also very active in engineering education and outreach programs that seek to broaden engineering talent in the STEM workforce. Dr. Bernadin may be reached at bernadin@eng.famu.fsu.edu

**Ali Hassan** is a Ph.D. student in Electrical Engineering at the FAMU-FSU College of Engineering, Department of Electrical and Computer Engineering. He earned his Bachelor of Science degree in Electrical Engineering from the Information Technology University, Lahore, Punjab. His current research focuses on using artificial intelligence for signal processing. He can be reached out at ali1.hassan@famu.edu